\documentclass[prl,aps,twocolumn, a4paper,longbibliography]{revtex4-2}

\usepackage[paperwidth=210mm,paperheight=297mm,centering,top=2cm,bottom=2.8cm,right=1.8cm,left=1.8cm]{geometry}

\pdfoutput=1

\date{\today}

\usepackage{xspace}
\usepackage{soul}

\newcommand{\um}{\upmu{\rm m}}
\newcommand{\nK}{\textrm{nK}}
\newcommand{\kB}{k_{\textrm{B}}}

\newcommand{\tuni}{\ensuremath{t_{\textrm {uni}}}\xspace}
\newcommand{\Tc}{T_{\textrm {c}}}

\newcommand{\upd}{\textrm{d}}
\newcommand{\potassium}{^{39}\textrm{K}}

\newcommand{\nzeq}{n_0^{\textrm{eq}}}

\newcommand{\ba}{\textbf{a}}
\newcommand{\bb}{\textbf{b}}
\newcommand{\bc}{\textbf{c}}
\newcommand{\bd}{\textbf{d}}
\newcommand{\be}{\textbf{e}}

\usepackage{float}

\usepackage{fourier}

\usepackage{color}
\usepackage{graphicx}
\usepackage{amsmath,amssymb}

\usepackage{times}
\usepackage{upgreek}
\usepackage{psfrag} 
\usepackage{latexsym} 
\usepackage{amstext}
\usepackage{amsxtra} 

\usepackage{lineno}

\usepackage{textcomp}
\usepackage{amsfonts}
\usepackage{graphicx}
\usepackage{bm}
\usepackage{color}
\usepackage{braket}
\usepackage{siunitx}
\usepackage[svgnames]{xcolor}
\usepackage[normalem]{ulem}

\definecolor{myColor}{rgb}{0.02,0.12,0.3}
\definecolor{myciteColor}{rgb}{0.39,0.7,0.89}
\usepackage[colorlinks=true,citecolor=myColor,linkcolor=myColor,urlcolor=myColor]{hyperref}

\def\be{\begin{equation}}
\def\ee{\end{equation}}

\makeatletter
\def\@fnsymbol#1{\ensuremath{\ifcase#1\or *\or \dagger\or \ddagger\or
   \mathsection\or \mathparagraph\or \|\or **\or \dagger\dagger
   \or \ddagger\ddagger \else\@ctrerr\fi}}
\makeatother

\usepackage{titlesec}

\begin{document} 

\title{
{A universal speed limit for spreading of coherence}
}

\author{Gevorg Martirosyan}
\email{gm572@cantab.ac.uk}
\author{Martin Gazo}
\author{Ji\v r\' i Etrych}
\author{Simon M.~Fischer}
\author{Sebastian~J.~Morris}
\author{Christopher J.~Ho}
\author{Christoph Eigen}
\author{Zoran Hadzibabic}
\email{zh10001@cam.ac.uk}
\affiliation{Cavendish Laboratory, University of Cambridge, J. J. Thomson Avenue, Cambridge CB3 0HE, United Kingdom}

\begin{abstract}
Discoveries of fundamental limits for the rates of physical processes, from the speed of light to the Lieb--Robinson bound for information propagation~\cite{Lieb:1972,Chen:2023},
often lead to breakthroughs in the understanding of the underlying physics.
Here we observe such a limit for a paradigmatic many-body phenomenon, the spreading of coherence during formation of a weakly interacting Bose--Einstein condensate~\cite{Snoke:1989,Stoof:1991,Svistunov:1991,Kagan:1992,Semikoz:1995,Kagan:1995ch,Damle:1996,Gardiner:1997,Miesner:1998b,Berloff:2002,Kohl:2002,Ritter:2007,Hugbart:2007,Smith:2012,Navon:2015,Proukakis:2024}. 
We study condensate formation in an isolated homogeneous atomic gas~\cite{Gaunt:2013,Navon:2021} that is initially far from equilibrium, in an incoherent low-energy state, and condenses as it relaxes towards equilibrium.
Tuning the inter-atomic interactions that drive condensation, we show that the spreading of coherence through the system is initially slower for weaker interactions, and faster for stronger ones, but always eventually reaches the same limit, where the square of the coherence length grows at a universal rate given by the ratio of Planck's constant and the particle mass{, or equivalently by the quantum of velocity circulation associated with a quantum vortex}. 
These observations are robust to changes in the initial state, the gas density, and the system size. Our results provide benchmarks for theories of universality far from equilibrium~{\cite{Kraichnan:1967,Kibble:1976,Svistunov:1995,Pomeau:1996,Bray:2002,Vinen:2002, Micha:2004,Berges:2008, Nazarenko:2011, Polkovnikov:2011, Eisert:2015,  Berges:2021,Barenghi:2023, Rosenhaus:2025}}, are relevant for quantum technologies that rely on large-scale coherence, and invite similar measurements in other systems.
\end{abstract}

\maketitle

 Understanding the dynamics of far-from-equilibrium many-body systems, including the emergence of long-range order in such systems, is an outstanding problem in physics, relevant from subnuclear to cosmological lengthscales~\cite{Kraichnan:1967,Kibble:1976,Svistunov:1995,Pomeau:1996,Vinen:2002,Bray:2002,Micha:2004,Berges:2008,Nazarenko:2011,Polkovnikov:2011,Eisert:2015,Berges:2021,Barenghi:2023,Rosenhaus:2025}.~{Conceptually, far-from-equilibrium relaxation and emergence of coherence have long been linked to decaying turbulence~\cite{Kraichnan:1967,Svistunov:1995,Berloff:2002,Micha:2004}, which often features self-similar scaling dynamics}. More recently, within the framework of nonthermal fixed points (NTFPs)~\cite{Berges:2008}, theorists have drawn analogies between such dynamics and the equilibrium properties of systems close to a continuous phase transition. Near the transition to an ordered state of matter, such as a superfluid or a ferromagnet, the system is scale invariant and its salient properties do not depend on the microscopic details~\cite{Chaikin:1995}.
Analogously, in the NTFP theory, far-from-equilibrium systems, including the early universe undergoing reheating~\cite{Micha:2004}, quark-gluon plasma in heavy-ion collisions~\cite{Berges:2014b}, quantum magnets~\cite{Bhattacharyya:2020}, and ultracold atomic gases~\cite{Nowak:2012, PineiroOrioli:2015, Berges:2015, Chantesana:2019,Groszek:2021,Chatrchyan:2021}, generically display dynamic (spatiotemporal) scaling, with scaling exponents that could define far-from-equilibrium universality classes.
Recently, far-from-equilibrium dynamic scaling was observed in several experiments with ultracold atoms, {in both isolated (relaxing)~\cite{Pruefer:2018,Erne:2018,Glidden:2021,Huh:2024, Madeira:2024,Gazo:2025,Liang:2025} and continuously driven~\cite{Galka:2022, Martirosyan:2024} systems}.

Here, we go beyond the elegant scaling properties of far-from-equilibrium relaxation~\cite{Pruefer:2018,Erne:2018,Glidden:2021, Huh:2024,Madeira:2024,Gazo:2025,Liang:2025}, to address the crucial question of how long it takes to establish long-range order.  We study this problem for the paradigmatic macroscopically coherent state, the weakly interacting Bose--Einstein condensate, which is also a textbook example of a superfluid.

\begin{figure*}[t!]
\centerline{\includegraphics[width=\textwidth]{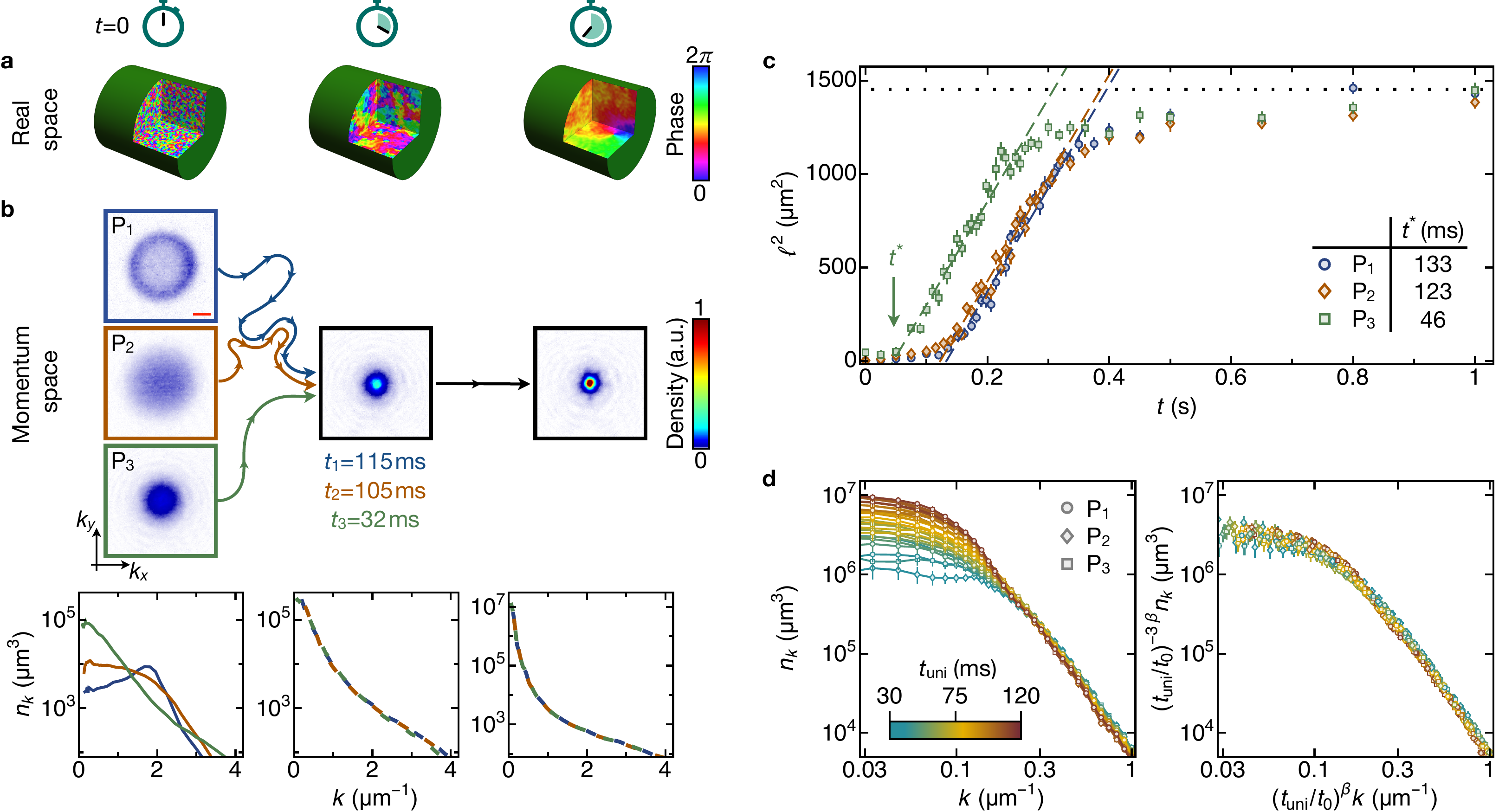}}
\caption{
\textbf{Universal coarsening of an isolated Bose gas.}
\ba, Real-space cartoon of coarsening. 
\bb, Momentum-space relaxation for different far-from-equilibrium initial states.
Our initial states P$_{1,2,3}$ (left column) have different momentum distributions $n_k$, but the same energy, so the gas always relaxes towards the same equilibrium state. For P$_{1,2,3}$, the system takes different times, $t_{1,2,3}$, to evolve to the same $n_k$ shown in the middle column, but from this point onwards it always evolves in the same way. 
The $n_k$ distributions are averages of at least $20$ measurements.
The red scale bar (top image) shows $1\,\um^{-1}$.
\bc, Growth of the coherence length, $\ell$ (see text). Plotting $\ell^2(t)$ reveals three stages of relaxation: (i) the non-universal initial dynamics, (ii) the scaling regime where $\ell^2$ grows linearly (dashed lines), as expected for the scaling exponent $\beta = 1/2$, and (iii) the breakdown of scaling at long times due to finite-size effects. The curves for P$_{1,2,3}$ are parallel, with the initial-state effects captured by the different time offsets $t^*$ (intercepts of the dashed lines). \bd, Dynamic scaling. In the scaling regime, the full low-$k$ distributions for all three initial states (left panel) can be collapsed onto the same curve (right panel) according to Eqs.~(\ref{eq:beta}, \ref{eq:dynamicScaling}) with $\beta=1/2$ and $t \rightarrow \tuni \equiv t - t^*$  (see also Extended Data Fig.~\ref{fig:S4}); $t_0=60\,\textrm{ms}$ is an arbitrary reference time. All error bars show standard errors of the measurements.
}
\label{fig:1}
\end{figure*}

We study condensate formation in an isolated homogeneous Bose gas~\cite{Eigen:2016}, trapped in a cylindrical optical box~\cite{Gaunt:2013, Navon:2021} such as sketched in Fig.~\ref{fig:1}\textbf{a} (Methods). The gas is prepared far from equilibrium and is initially incoherent, but has very low energy and condenses as it relaxes towards equilibrium under the influence of inter-atomic interactions, characterised by the $s$-wave scattering length $a$. 
As illustrated in Fig.~\ref{fig:1}\textbf{a}, in a homogeneous system the (global) condensate grows through coarsening~\cite{Bray:2002}, the local spreading of coherence. 
This coarsening is quantified by the growth of the coherence length $\ell$, { over which the first-order correlation function $g_1(r)$ decays, and corresponds to narrowing of the momentum distribution $n_k (k)$ (where $k$ is the wavenumber), which is related to $g_1(r)$ by the Fourier transform.}

\begin{figure*}[t!]
\centerline{\includegraphics[width=\textwidth]{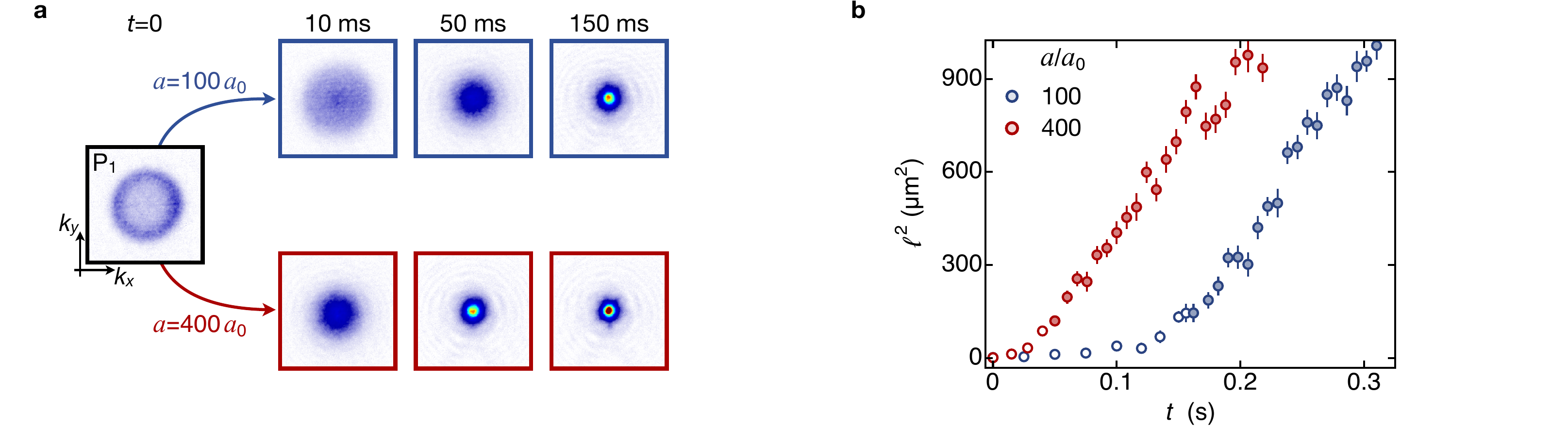}}
\caption{
\textbf{Spreading of coherence for different interaction strengths.}
\textbf{a}, Gas evolution for two different scattering lengths $a$, starting in the same initial state P$_1$. For stronger interactions, the condensate (the peak at $k=0$) emerges sooner.
\textbf{b}, However, plotting $\ell^2 (t)$ reveals that the interaction strength affects only the non-universal initial dynamics (open symbols), while the linear growth of $\ell^2$ in the universal coarsening regime (solid symbols) is the same for both $a$ values.
}
\label{fig:2}
\end{figure*}

Our experiments are performed with $\potassium$ atoms, and we tune $a$ using a Feshbach resonance, exploring coarsening for $a = (50-400)\,a_0$, where $a_0$ is the Bohr radius. 
Our cylindrical box has radius $R=21(2)\,\um$, length $L=40(4)\um$, and volume $V = 55(12) \times 10^3\,\um^3$. Our gas density $n=5.4(1.2)\,\um^{-3}$ corresponds (assuming ideal-gas thermodynamics) to the critical temperature for condensation $\Tc = 127(19)\,\nK$. 
The {kinetic} energy per particle in our initial incoherent states is $\varepsilon = \kB \times 20(2)\,\nK$ (where $\kB$ is the Boltzmann constant), corresponding to a large equilibrium condensed fraction $\eta = 0.61(4)$ (see Extended Data Figs.~\ref{fig:S1} and~\ref{fig:S2}).
{During coarsening the total particle number, $N \approx 3\times 10^5$, is essentially constant (see Methods),} and the gas is always weakly interacting in the sense that $na^3< 10^{-4}$.
We measure $n_k$ by absorption imaging (along the $z$ direction) after time-of-flight expansion, performing the inverse-Abel transform on the line-of-sight integrated distributions; just for the images shown in Figs.~\ref{fig:1}{\bb} and ~\ref{fig:2}{\ba} we instead image only slices of the cloud~\cite{Andrews:1997a} corresponding to $k_z\approx 0$ (Methods).
We normalise $n_k$ such that $\int n_k \, 4\pi k^2 \upd k = N$.

We first show that, while the short-time relaxation dynamics inevitably depend on the details of the initial state, the long-time relaxation does not (Fig.~\ref{fig:1}{\bb}).
For this purpose, we engineer three different far-from-equilibrium states, starting with a quasi-pure condensate and using a time-varying force to perturb the cloud (see Extended Data Fig.~\ref{fig:S1}). Our initial states P$_{1,2,3}$ have different $n_k$ (see left column), but the same $\varepsilon$. At time $t=0$ the gas is non-interacting and we then initiate relaxation by switching $a$ to $100\,a_0$. Starting from an initial state, $n_k(k)$ follows some trajectory (represented by the wiggly coloured lines) in the space of functions with the same $N$ and $\varepsilon$. The middle column shows that, still far from equilibrium, these trajectories converge to the same $n_k$. The time that the gas takes to evolve to this $n_k$ depends on the initial state ($t_{i=1,2,3}$ for P$_{i = 1,2,3}$), but further evolution from this $n_k$ is the same for all initial states; note that the state trajectories for P$_1$ and P$_2$ merge before merging with the P$_3$ one, but we just show an $n_k$ where all three have converged.

{The long-time relaxation of a low-energy Bose fluid was theoretically studied in different frameworks. In Ref.~\cite{Svistunov:1995}, this problem was studied for an incompressible superfluid, where the spreading of coherence is associated with the decay of a random tangle of quantized vortices (variously known as the Kibble's vortex tangle~\cite{Kibble:1976}, superfluid turbulence~\cite{Feynman:1955}, and Vinen turbulence~\cite{Vinen:2002,Barenghi:2023}) and $\ell$ is set by the typical distance between the vortex lines. For $\ell \gg \xi$, where $\xi$ is the size of the vortex core, the prediction is that
\begin{equation}
\label{eq:Svistunov}
    \frac{\upd \ell}{\upd t} \propto \frac{\ln{[A \ell/\xi]}} {\ell} \, ,
\end{equation}
where $A$ is a dimensionless constant. 
On the other hand, for coarsening of wave excitations, corresponding to a compressible-fluid flow, approximate kinetic equations give
\begin{equation}
\label{eq:beta}
    \ell(t) \propto t^{\beta} \, ,
\end{equation}
with $\beta = 1/2$~\cite{Kraichnan:1967,PineiroOrioli:2015,Chantesana:2019,Rosenhaus:2025}.
In our weakly interacting gas, with the vortex-core size set by the healing length $\xi = 1/\sqrt{8\pi na}$, both vortices and waves could play a significant role. Note, however, that the predictions of Eqs.~(\ref{eq:Svistunov}) and (\ref{eq:beta}) differ only in a logarithmic correction. Moreover, our measurements of $n_k$ probe $\ell$ independently of what type of excitations are dominant and limit its value.

For coarsening fully characterised by the growth of $\ell$~\cite{Bray:2002}, the low-$k$ momentum distribution exhibits self-similar evolution:
\begin{equation}
\label{eq:dynamicScaling}
    n_k(k,t)= \ell^d(t) \,  f\left[k\ell(t)\right] \, ,
\end{equation}
where $d=3$ is the system dimensionality and $f$ is a dimensionless scaling function. In this case $\ell(t)$ is $\propto n_0^{1/3}(t)$, where $n_0\equiv n_k(k=0)$, and the condensate occupation is $N_0 = (2\pi)^3 n_0/V$. This scaling does not define the absolute value of $\ell$, and here we define it so that in equilibrium $\ell^3$ is equal to the system volume; to convert the observed $n_0$ values to $\ell$ we also take into account the finite $k$-space resolution of our time-of-flight measurements (Methods). 
We observe good agreement with Eqs.~(\ref{eq:beta}, \ref{eq:dynamicScaling}), without the need to invoke the logarithmic correction in Eq.~(\ref{eq:Svistunov}), but this does not exclude its relevance on much larger lengthscales; also note that the importance of this correction depends on the unknown value of $A$.
}

To compare our data to Eq.~(\ref{eq:beta}), we first note that it is formally valid only for $t, \ell \rightarrow \infty$. However, one can observe the same scaling for finite $t$ and $\ell$ by absorbing all the effects of the non-universal initial dynamics into time offsets such as seen in Fig.~\ref{fig:1}{\bb}, {\it i.e.},  by shifting $t\rightarrow t - t^*$, where $t^*$ depends on the initial state~\cite{Gazo:2025} (see also ~\cite{Heller:2024b}).

For $\beta = 1/2$, in the scaling regime $\ell \propto (t -t^*)^{1/2}$, and in Fig.~\ref{fig:1}{\bc} we plot $\ell^2(t)$. This reveals both  the scaling regime where $\ell^2$ grows linearly, at a rate that does not depend on the initial state (dashed lines), and the offsets $t^*$ for P$_{1,2,3}$; see also Extended Data Fig.~\ref{fig:S3}.
As the system approaches equilibrium, for $\ell^2 \gtrsim 900\,\um^2 \approx (V/2)^{2/3}$ the scaling breaks down due to finite-size effects; from hereon we focus on the regime before this breakdown.

In Fig.~\ref{fig:1}{\bd}, using the $t^*$ values from Fig.~\ref{fig:1}{\bc}, we show that in the scaling regime the low-$k$ momentum distributions for all three initial states can indeed be collapsed onto the same universal curve {according to Eq.~(\ref{eq:dynamicScaling}) with $\ell\propto (t-t^*)^{1/2}$ (see also Extended Data Fig.~\ref{fig:S4})}.

We now turn to varying the strength of the interactions that drive the coarsening, and study how this affects the coarsening rate.
For $\ell\propto (t-t^*)^{\beta}$, {we define the `speed of coarsening' as} $D\equiv\upd \ell^{1/\beta}/\upd t$, which is (for any $\beta$) time-invariant in the scaling regime and does not depend on the non-universal $t^*$.
For a low-energy Bose gas described by the Gross--Pitaevskii equation, the interaction-set units of length and time are $\xi$ and $t_{\xi} \equiv m\xi^2/\hbar$, where $\hbar$ is the reduced Planck's constant and $m$ the atom mass. Hence, on dimensional grounds  $\ell/\xi \propto [(t - t^*)/t_{\xi}]^{\beta}$ and {we get}
\begin{equation}
\label{eq:Ddim}
    D \ \propto \ \frac{\xi^{1/\beta}}{t_{\xi}} \ \propto \ \frac{\hbar}{m} \, (na)^{1 - 1/(2\beta)} \, .
\end{equation}
Specifically for $\beta = 1/2$, as observed in Fig.~\ref{fig:1}, this has a counter-intuitive implication that $D = \upd \ell^2 / \upd t \sim \hbar/m$ does not depend on the interaction strength {$na$} (see also~\cite{Kraichnan:1967,Pomeau:1996}). 
{In the hydrodynamic theory of Ref.~\cite{Svistunov:1995}, the same result emerges if one neglects the logarithmic correction in Eq.~(\ref{eq:Svistunov}) (which depends on the interactions through $\xi$), because then $\upd \ell^2 / \upd t$ can only be set by the quantum of velocity circulation associated with a quantum vortex, $\kappa = 2\pi \hbar/m$.
}

\begin{figure}[t!]
\centerline{\includegraphics[width=\columnwidth]{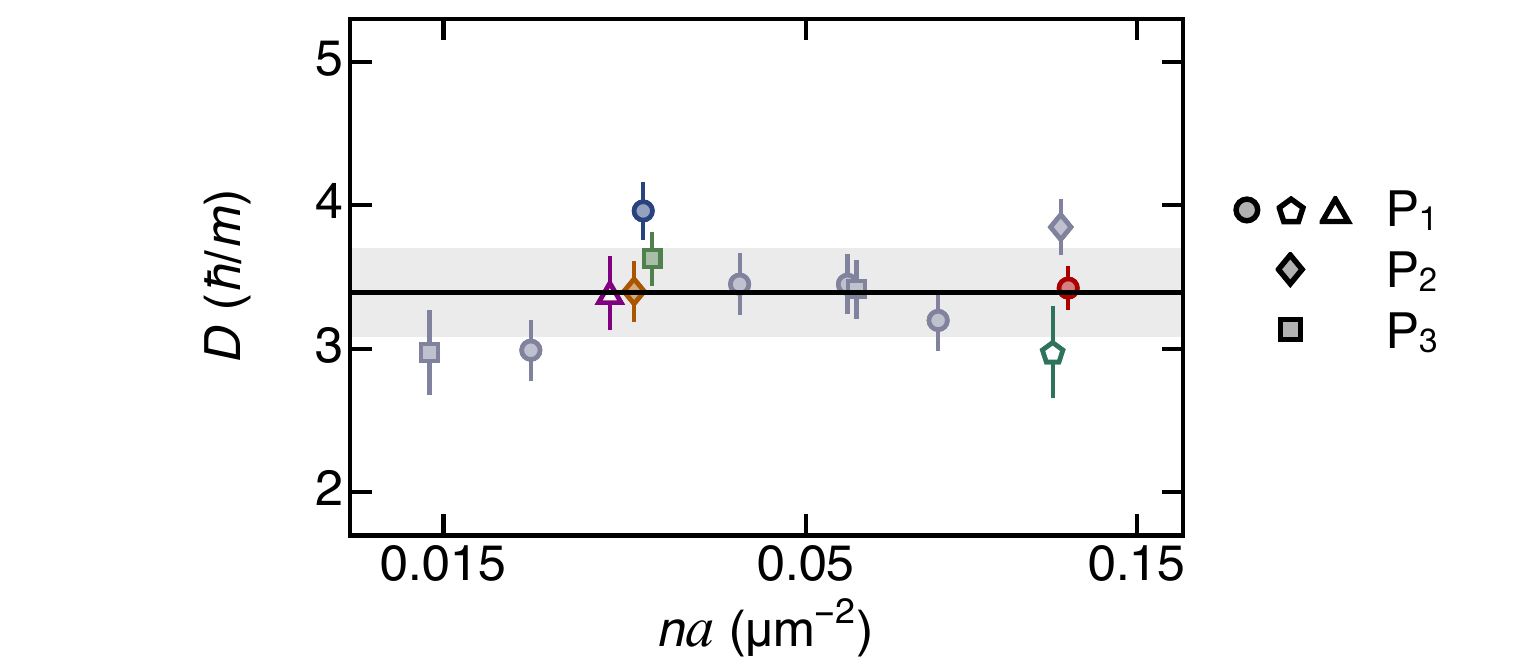}}
\caption{
\textbf{Universality of the coarsening speed $D$.}
Our measurements for various interaction strengths show no systematic variation of $D = \upd \ell^2 / \upd t$ (in the scaling regime), and give a combined estimate $D=3.4(3)\,\hbar/m$ (solid line and shading). 
{The four data sets shown in Fig.~\ref{fig:1}{\bc} or Fig.~\ref{fig:2}{\bb} are represented here by the corresponding coloured symbols.}
The purple triangle indicates a measurement where we reduced the gas density $n$ by a factor of $4.2$, which reduces $\Tc \propto n^{2/3}$ and the equilibrium condensed fraction (from about $60$\% to about $30$\%), but $D$ is not affected.
The green pentagon indicates a measurement where we instead reduced the system volume $V$ by a factor of $3.5$; in this case $\ell^2$ saturates at a correspondingly lower value ($\propto V^{2/3}$), but again $D$ is not affected.  {For details on the data taken with reduced $n$ or $V$} see Extended Data Fig.~\ref{fig:S5}. The data points corresponding to the same $na$ values are slightly offset horizontally for visual clarity.
}
\label{fig:3}
\end{figure}

\begin{figure*}[t!]
\centerline{\includegraphics[width=\textwidth]{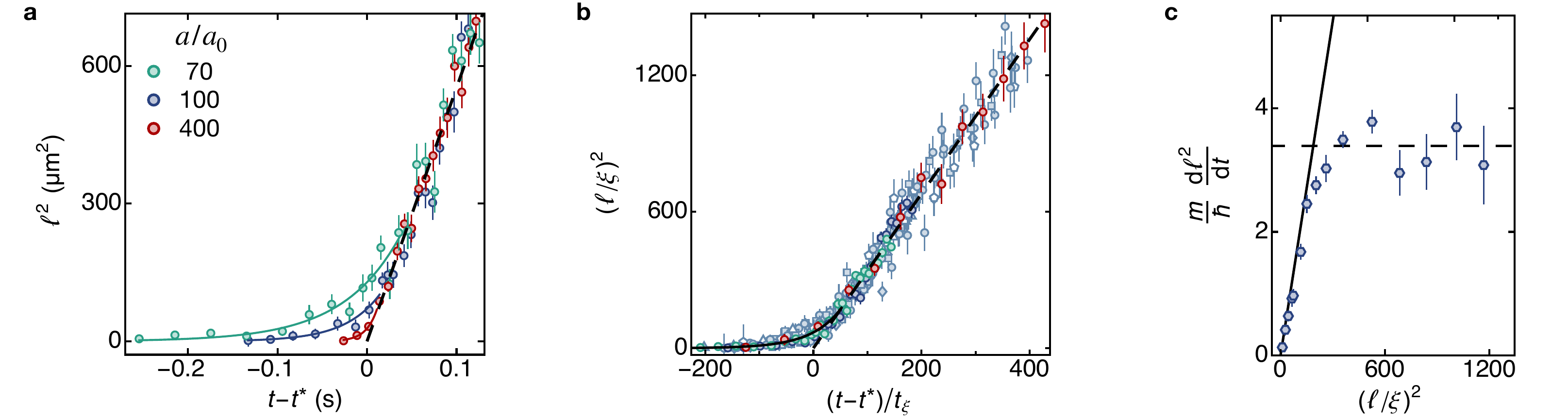}}
\caption{
{
\textbf{Reaching the coarsening speed limit.}
{\ba}, The growth of $\ell^2$ for the same initial state (P$_1$) and gas density, but different $a$. All curves start at $t=0$ and plotting them versus $t-t^*$ (with $a$-dependent $t^*$) collapses them in the universal scaling regime. The dashed line shows $\ell^2 = D (t - t^*)$. For weaker interactions the system approaches this speed limit slower, and reaches it at a larger $\ell$; the solid lines show exponential fits to the early-time data.
{\bb}, 
When expressed in the interactions-set units of length, $\xi = 1/\sqrt{8\pi na}$, and time, $t_{\xi} \equiv m\xi^2/\hbar$, all our data 
for different P$_{i}$, $a$, $V$, and $n$ (see Fig.~\ref{fig:3}) collapse onto a single curve, meaning that the speed limit is always reached at the same $\ell/\xi$.
The solid line shows exponential growth with a time constant $\tau = 56 \, t_{\xi}$ and the dashed line has slope $mD/\hbar =3.4$. {\bc},~Numerically differentiating the data in {\bb}, we eliminate the non-universal $t^*$ and show as a function of $(\ell/\xi)^2$ how $\upd \ell^2 / \upd t$ approaches the universal $D = 3.4\, \hbar/m$ and then stops growing; the solid and dashed lines are the same functions as in {\bb}. 
For weaker interactions (larger $\xi$), observing this speed limit requires a larger physical system.}
}
\label{fig:4}
\end{figure*}

In Fig.~\ref{fig:2} we show data for coarsening at $a=100\,a_0$ and $400\,a_0$, starting from the same initial state P$_1$; the $100\,a_0$ data here is the same as in Fig.~\ref{fig:1}.
The images in Fig.~\ref{fig:2}{\ba}  show what one intuitively expects -- for larger $a$ the condensate emerges sooner. However, in Fig.~\ref{fig:2}{\bb}, plotting $\ell^2 (t)$ reveals that the effects of the interaction strength are, just like the initial-state effects, confined to the difference in the non-universal $t^*$ -- in the scaling regime (solid symbols), the rate of the linear growth of $\ell^2$ is essentially the same for both $a$.

We performed such measurements for various scattering lengths, and also varied the gas density and system size. In each case we fit the slope $ D = \upd \ell^2 / \upd t$ in the scaling regime, and summarise our results in Fig.~\ref{fig:3}.
We observe no systematic variation of $D$ {with the interaction strength $na$} and get a combined estimate {$D=3.4(3)\,\hbar/m$; here the error in $D$ is purely statistical, while the uncertainty in our box volume leads to a systematic error of $\pm \, 0.5\,\hbar/m$ (Methods)}.

Finally, in Fig.~\ref{fig:4} we study how the system approaches the long-time coarsening speed $D$, which also reconciles our observations with the finite-system intuition (and experience) that for stronger interactions condensates form faster, and for $a\rightarrow 0$ they do not form at all.

As we illustrate in Fig.~\ref{fig:4}{\ba} for fixed $n$, the same initial state (P$_1$), and three values of $a$, for weaker interactions the gas takes longer to join the universal scaling trajectory, and joins it at a larger value of $\ell$. Here all curves start at $t=0$ and plotting versus $t-t^*$ collapses them in the scaling regime; the dashed line shows $\ell^2 = D(t-t^*)$.

In Fig.~\ref{fig:4}{\bb} we show that expressing all our data for different P$_{i}$, $a$,  $V$, and $n$ (see Fig.~\ref{fig:3}) in terms of $\xi$ and $t_{\xi}$ collapses them onto a single curve. Here, the dashed line has slope $mD/\hbar = 3.4$ and the solid line that captures the approach to the scaling regime is an exponential with a time constant $\tau = 56 \,t_{\xi}$ (see also Extended Data Fig.~\ref{fig:S6}). Numerically differentiating these data, and thus eliminating the non-universal $t^*$, in Fig.~\ref{fig:4}{\bc} we show how $\upd \ell^2/\upd t$ approaches the universal $D=3.4\,\hbar/m$ (dashed line) as a function of $(\ell/\xi)^2$.

The dimensionless results in Figs.~\ref{fig:4}{\bb} and \ref{fig:4}{\bc} imply that for any interaction strength the system would eventually, for $\ell \gg \xi$, exhibit the same coarsening speed $D$.
However, the system size required to observe this is larger for larger $\xi$ (smaller $na$), and diverges for $na\rightarrow 0$. As we discuss in Methods and Extended Data Fig.~\ref{fig:S7}, previous experiments on the emergence of extended coherence during far-from-equilibrium condensation, in both harmonic~\cite{Ritter:2007} and box~\cite{Glidden:2021} traps, were not in the universal-speed regime; 
consequently, for tunable interactions in Ref.~\cite{Glidden:2021}, the observed relaxation time was $\propto 1/a$.

{Our results should be relevant across many fields, from cold atoms and the conventional low-temperature physics~\cite{Vinen:2002,Barenghi:2023} to cosmology~\cite{Kibble:1976,Micha:2004} and high-energy physics~\cite{Berges:2015}.
The fact that $\hbar$ and $m$ appear in $D$ only through their ratio, or the quantum of circulation $\kappa$, implies that the results are also applicable to systems where the underlying physics is not quantum~\cite{Svistunov:2015}. They should also be relevant for benchmarking the theories of ultrarelativistic systems, for which $\beta = 1/2$ is also predicted~\cite{PineiroOrioli:2015}, but in that case the effective mass, $m_\textrm{eff}\sim\hbar/(\xi c)$ (where $c$ is the speed of light), and hence $D \sim \xi c$,  depend on the interactions.

The value of $D=3.4\, \hbar/m$, equal to $5.5\,\um^2/{\rm ms}$ for $^{39}$K, has curious implications for the emergence of coherence on truly macroscopic lengthscales. For example, for coherence to spread via coarsening over $1\,$cm would require hours, and even possible logarithmic corrections cannot change this conclusion significantly. However, an interesting question is whether this speed limit can be broken by a fundamentally different preparation protocol, for example by melting a Mott insulator (through a quantum phase transition) to obtain a superfluid with long-range coherence.}

In the future, it would be interesting to {disentangle the roles of waves and vortices during coarsening, by directly imaging the latter, to search for possible logarithmic corrections to the coarsening speed}, and to perform similar measurements for fermionic superfluids and gases with long-range interactions.

We thank Andrey Karailiev, Maciej Ga{\l}ka, Timon Hilker, Mikhail Lukin, Robert Smith, Tobias Donner, Gregory Eyink, and Vladimir Rosenhaus for useful discussions and comments on the manuscript.
This work was supported by EPSRC [Grants No.~EP/P009565/1 and EP/Y01510X/1], ERC (UniFlat), and STFC [Grants No.~ST/T006056/1 and No.~ST/Y004469/1].
C.~E. acknowledges support from Jesus College (Cambridge). Z.~H. acknowledges support from the Royal Society Wolfson Fellowship.

{\bf Author contributions} \quad G.M. led the project and collected the data. G.M., M.G., J.E., C.E., and Z.H. conceptualised the project. All authors contributed to the experimental setup, data analysis, and writing of the manuscript.

{\bf Competing interests}\quad The authors declare no competing interests.

%


\clearpage
\setcounter{figure}{0} 
\setcounter{equation}{0}

\renewcommand\theequation{S\arabic{equation}} 
\renewcommand\thefigure{\arabic{figure}} 
\renewcommand{\figurename}[1]{Extended Data Fig.~}

\titlespacing\subsection{0pt}{11pt}{11pt}

\section{METHODS}

\subsection{Optical box trap}

Our cylindrical optical box is made of $532\,$nm light {and has a trap depth of $\approx\kB \times 300\,{\rm nK}$}. Our standard box of volume $V = 55(12) \times 10^3\,\um^3$, used for most measurements, has radius $R= 21(2) \,\um$ and length $L = 40(4)\, \um$. 
For the additional measurement with the smaller $V = 16(3) \times 10^3\,\um^3$, we use a box with $R = 14(1) \,\um$ and $L = 26(3)\, \um$. Note that in both cases $V^{1/3} \approx 2R \approx L$.

\subsection{Measuring the momentum distribution $n_k(k)$}
\vspace{0.5em}
Our experiments are performed with the lowest hyperfine ground state of $^{39}$K. We measure $n_k(k)$ after a time-of-flight (ToF) expansion {(at the start of which we set $a \rightarrow 0$)}, first optically pumping the atoms to the highest hyperfine ground state and then imaging them. To deduce $n_k$ values that vary over 6 orders of magnitude, we combine measurements with ToF duration in the range $(16-120)\,{\rm ms}$; the longest ToF gives the best $k$-space resolution, relevant for low $k$, while shorter ones give better signal-to-noise at large $k$. Even for our longest ToF, the optical density (OD) at $k\approx0$ is high for clouds with a high condensed fraction, so to deduce high column densities while keeping ${\rm OD}\lesssim 2$, we pump a variable fraction of atoms (down to $\approx3\,\%$) into the imaging state~\cite{Ramanathan:2012}.

{Just for the images shown in Fig.~\ref{fig:1}\textbf{b} and Fig.~\ref{fig:2}\textbf{a}, and in Extended Data Fig.~\ref{fig:S1}, we image only slices of the cloud~\cite{Andrews:1997a} corresponding approximately to $k_z\in [-0.1,\, 0.1]\um^{-1}$, using a thin sheet beam to pump the atoms to the imaging state. Note that the thickness of our sheet pumping beam is not perfectly uniform, resulting in a slight left-right asymmetry in the OD; this is most visible in the image of the P$_1$ state, where the asymmetry is the largest, about $10\%$.}

\subsection{Initial-state preparation}

\begin{figure}[t!]
\centerline{\includegraphics[width=\columnwidth]{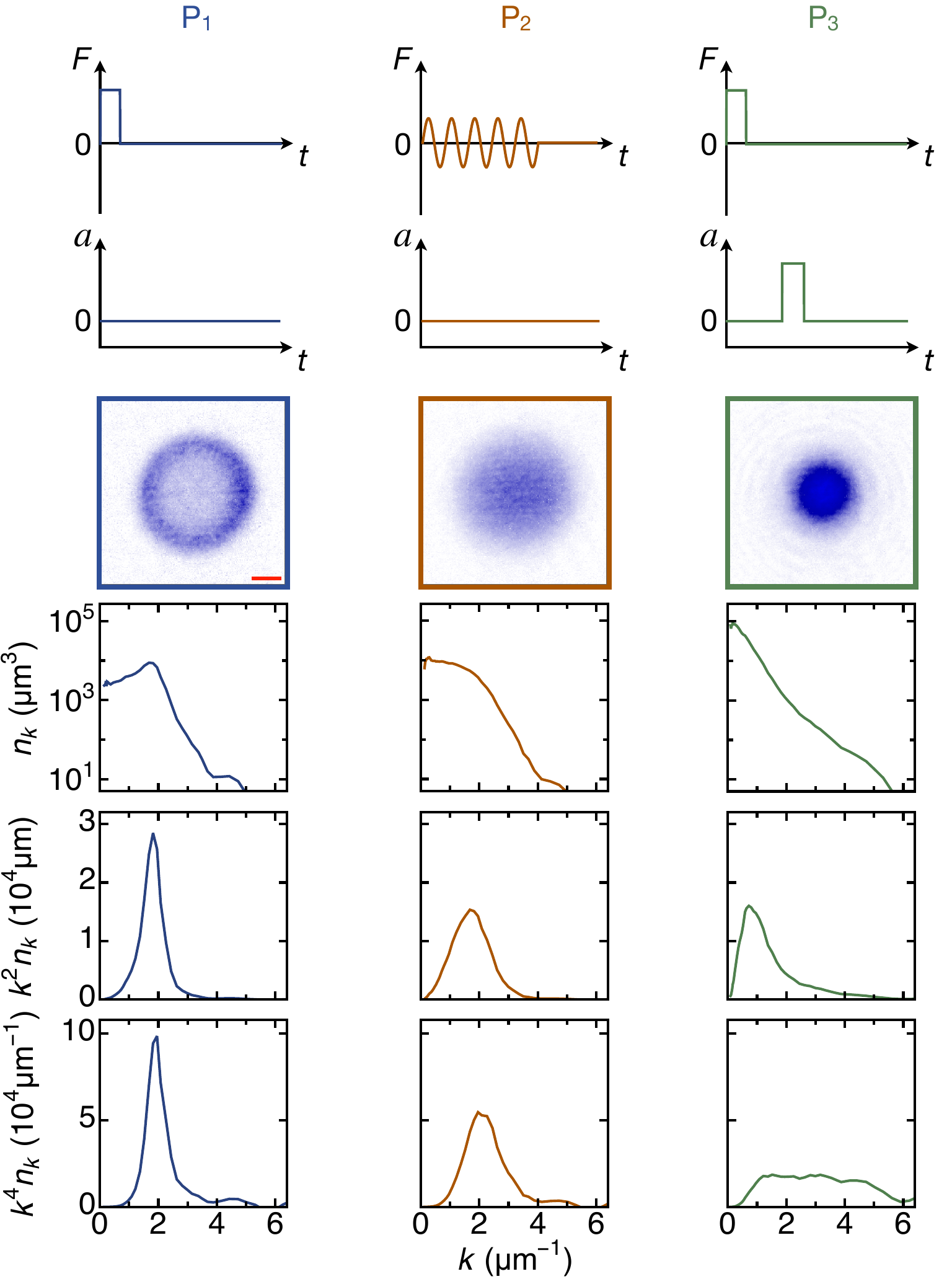}}
\caption{\textbf{Initial-state preparation.} We prepare our initial states using a combination of time-dependent force $F$ and interaction pulsing, as shown in the top two rows. The forcing pulses for P$_1$ and P$_3$ are $8\,\rm{ms}$ long and have strength $F_0=\kB \times 1.5\,{\rm nK}/\um$. The sinusoidal force for P$_2$ has amplitude $F_0=\kB \times 0.3\,{\rm nK}/\um$ and angular frequency $\omega = 2\pi \times 10\,$Hz, and is applied for $1\,$s. For P$_3$ the interactions are pulsed to $a=400\,a_0$ for $30\,$ms. After the end of the preparation sequence we wait for $500\,$ms at $a=0$ for $n_k$ to become isotropic. The images (same as in Fig.~\ref{fig:1}{\bb}) show the $k_x - k_y$ distributions for $k_z\approx 0$ just before we turn on interactions to initiate the relaxation; the red scale bar in the left image shows $1\,\um^{-1}$. The bottom three rows show different moments of the corresponding momentum distributions. The integrals of $k^2 n_k$ and $k^4 n_k$ are, respectively, proportional to the total atom number and energy, which are the same for all three states; the energy per particle is $\varepsilon = \kB \times 20(2)\,\nK$. 
}
\label{fig:S1}
\end{figure}

To prepare our initial states P$_{1,2,3}$, we start with a weakly interacting quasi-pure condensate with $N=\num{3.0(1)e5}$ atoms~\cite{Eigen:2016}, turn off the interactions ($a\rightarrow 0$) using the Feshbach resonance at $ 402.7\,\mathrm{G}$~\cite{Etrych:2023}, and perturb the cloud using a combination of time-dependent forcing {(using a magnetic field gradient)} and interaction pulsing, as outlined in Extended Data Fig.~\ref{fig:S1}. 
In a clean cylindrically symmetric potential, our forcing would result in anisotropic momentum distributions. However, weak disorder that is naturally present in our trap~\cite{Martirosyan:2024} couples excitations along different directions, so simply waiting for $500\,\mathrm{ms}$ at $a=0$ always results in states with isotropic, but far from equilibrium $n_k$.

\begin{figure*}[t!]
\centerline{\includegraphics[width=\textwidth]{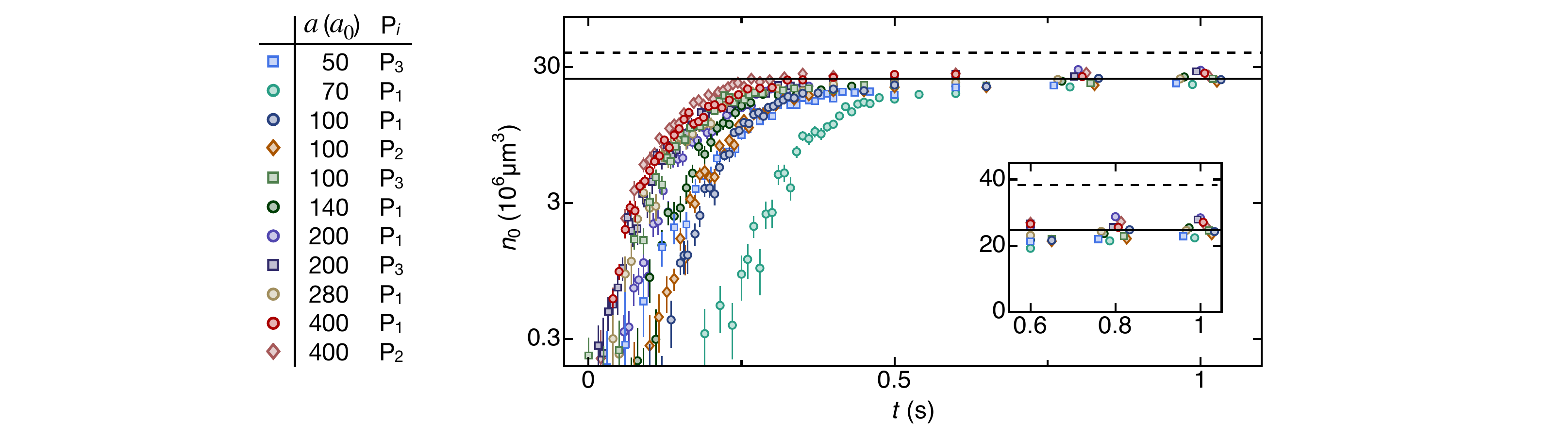}}
\caption{
{\textbf{Equilibrium $n_0$ and $\eta$. 
} We plot the observed $n_0(t)$ for all our $11$ data sets taken with $V=55(12)\times 10^3 \um^3$ and $n=5.4(1.2)\,\um^{-3}$. The solid line shows the long-time saturation value $\bar{n}_0 = 25(2) \times 10^6 \um^3$, obtained by averaging all the measurements for the two longest times ($0.8\,$s and $1.0\,$s for all data sets; the data points for these two times are slightly offset horizontally for visual clarity). The dashed line shows the value of $n_0$ observed for a quasi-pure BEC, $n_0^{\rm BEC}=39(2)\times 10^6 \um^3$, which gives $\bar{n}_0/n_0^{\rm BEC} = 0.64(6)$. The inset shows the same data on a linear scale, focusing on long times.}}
\label{fig:S2}
\end{figure*}

The forcing parameters are chosen such that P$_{1,2,3}$ all have the same energy per particle, $\varepsilon = \kB \times 20(2)\,$nK; for our $11$ data sets for which $V$ and $n$ are also the same, this corresponds to the same equilibrium state {(see Extended Data Fig.~\ref{fig:S2})}.

{\subsection{Deducing $\ell$ from $n_0$}
In this section, we explain how we deduce $\ell$ from the measured $n_0$, taking into account the finite $k$-space resolution of our time-of-flight measurements.

In Extended Data Fig.~\ref{fig:S2}, we show our $11$ main data sets, taken with $V=55(12)\times 10^3\,\um^3$ and $n= 5.4(1.2)\,\um^{-3}$. At long times, $n_0$ always saturates to $\bar{n}_0 = 25(2) \times 10^6 \,\um^3$ (solid line), which means that the equilibrium condensed fraction $\eta$ is always (approximately) the same. For comparison, for a quasi-pure equilibrium BEC ($\eta > 0.9$), which was never perturbed in the way shown in Extended Data Fig.~\ref{fig:S1}, we observe $n_0^{\rm BEC}=39(2)\times 10^6 \um^3$ (dashed line).

The ratio $\bar{n}_0/n_0^{\rm BEC} = 0.64(6)$ is consistent with our thermodynamic estimate $\eta=0.61(4)$ for $\varepsilon = \kB \times 20(2)\,$nK. However, in both cases, the observed equilibrium $n_0$ is, due to the finite $k$-space resolution, lower than the theoretical $n_0^{\rm th}=N_0 V/(2\pi)^3$ by a factor of $\zeta = 1.7(4)$, with the error in $\zeta$ dominated by the uncertainty in the box volume; 
note that $\zeta^{1/3} = 1.2(1)$ means that the narrowest observed $n_k$ distributions (corresponding to $\ell^3 = V$) are $20(10)\%$ broader than the Heisenberg limit set by the box size~\cite{Gotlibovych:2014}. }

\begin{figure}[b!]
\centerline{\includegraphics[width=\columnwidth]{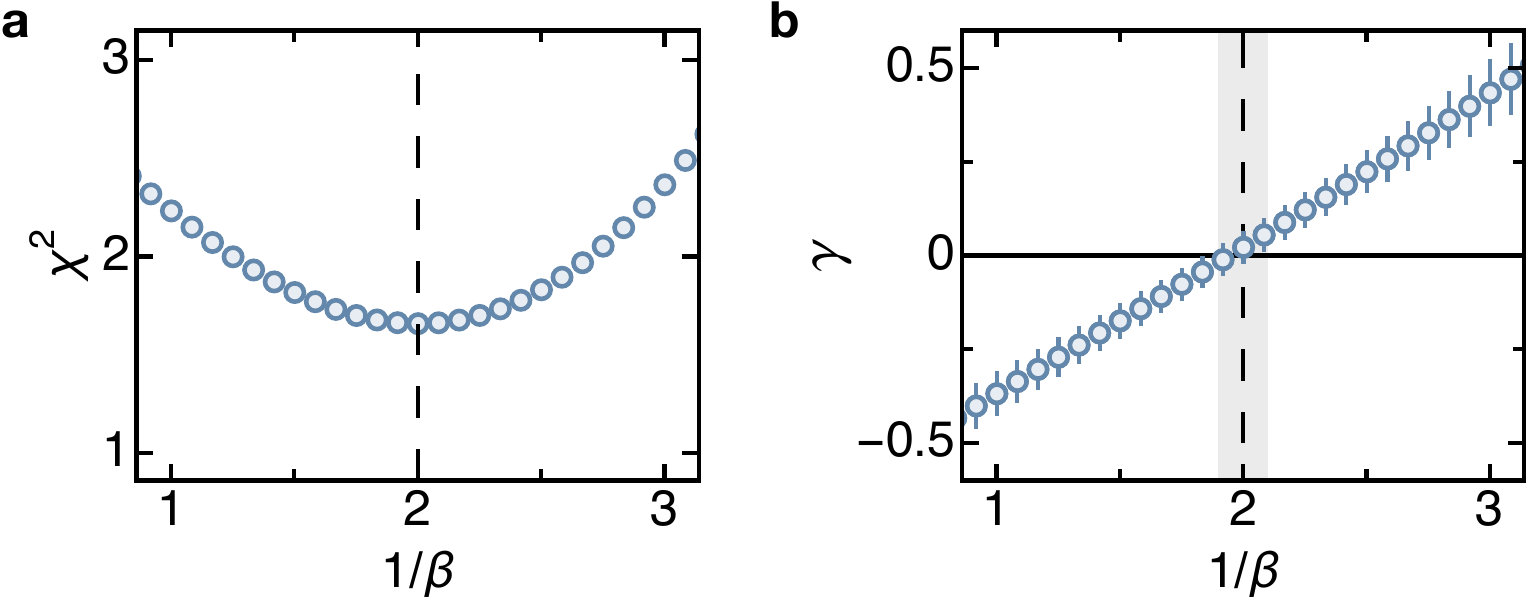}}
\caption{\textbf{Determining $\beta$.}
\textbf{a}, The $\chi^2$ values (combined for all $13$ data sets summarised in Fig.~\ref{fig:3}) for the linear long-time fits of $\ell^{1/\beta}(t)$, showing a minimum at $1/\beta \approx 2$.
\textbf{b}, The exponent $\gamma$ quantifying the self-consistency of the dependence of $\upd \ell^{1/\beta}/\upd t$ on $na$ for various $1/\beta$ (see text). Self-consistency requires $\gamma =0$, which we observe for $\beta=0.50(3)$ (grey shading).
}
\label{fig:S3}
\end{figure}

To model the effect of the finite $k$-space resolution, we assume that the true width of the momentum distribution ($\propto 1/\ell$) and the width of the resolution point spread function add in quadrature, so the apparent coherence length $\ell^{\prime}$ is related to $\ell$ by
\begin{equation}
1/{\ell^{\prime}}^2=1/\ell^2+1/\ell_0^2 \, ,
\end{equation}
where $\ell_0$ is set by our resolution, and we use our equilibrium measurements to calibrate $\ell_0=V^{1/3}/\sqrt{\zeta^{2/3}-1}=58(10) \um$. Hence, to deduce $\ell$ from the observed $n_0$, we first calculate $\ell^{\prime}=(n_0/\nzeq)^{1/3}V^{1/3}$ using $\nzeq =\zeta \bar{n}_0$, and then calculate $\ell^2={\ell^\prime}^2/[1-{(\ell^\prime/\ell_0)}^2]$. Note that if we numerically model the effect of the finite resolution as a convolution of the true $n_k$ with a Gaussian point spread function of width $\sigma_k$, our $\zeta=1.7$ corresponds to $\sigma_k \approx 0.04\, \um^{-1}$. 

For the values of $\ell$ reported in the main paper we use the central values for $V$ and $\ell_0$, and the error in $D = 3.4(3)\,\hbar/m$ is purely statistical, arising from the data scatter. The correlated errors in $V$ and $\ell_0$ correspond to a systematic error in $D$ of $\pm \, 0.5\, \hbar/m$.

\begin{figure}[b!]
\centerline{\includegraphics[width=\columnwidth]{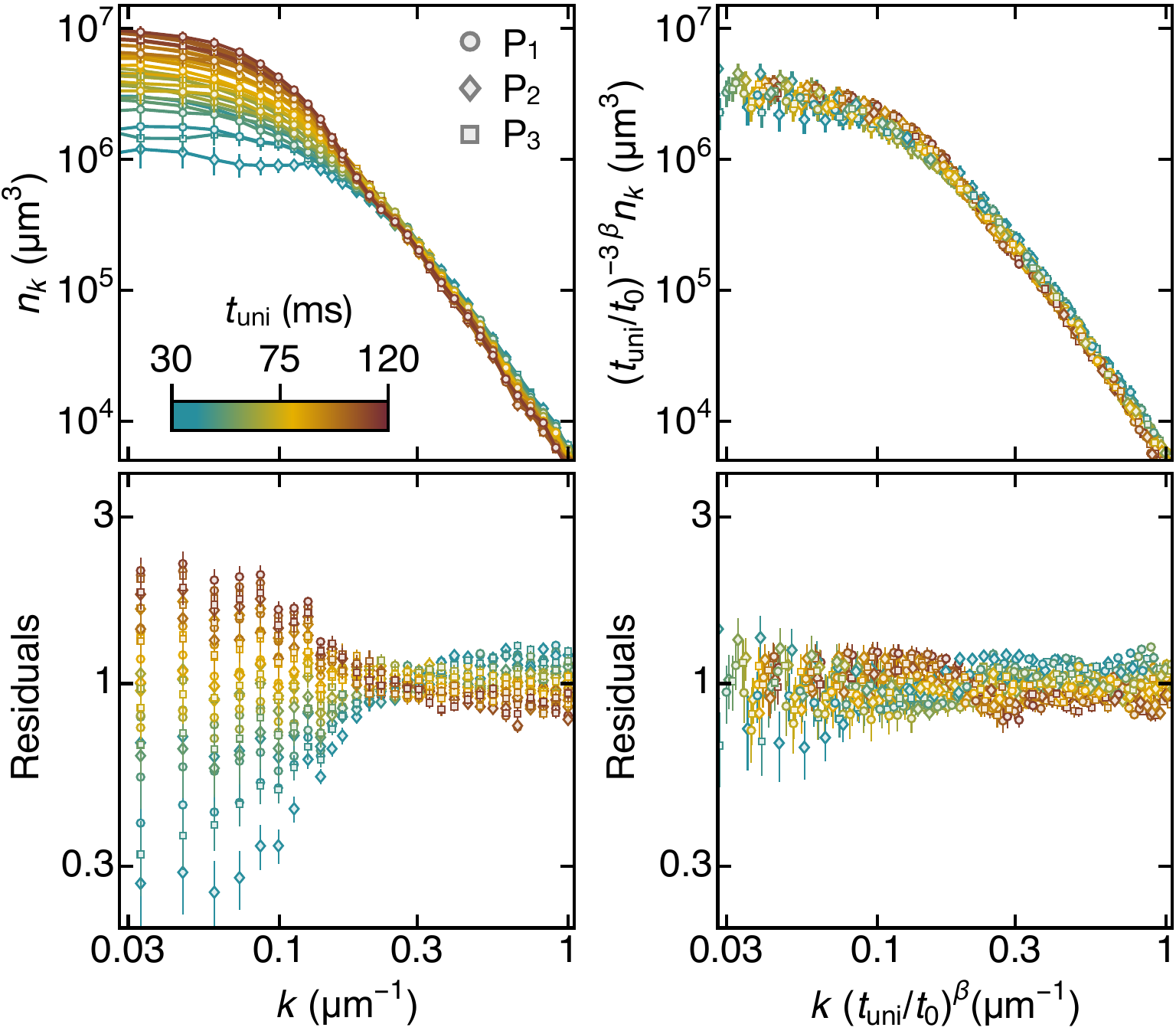}}
\caption{\textbf{Dynamic scaling residuals.}
The top panels reproduce Fig.~\ref{fig:1}{\bd}, and the bottom panels show the corresponding log-space spread of the data.
}
\label{fig:S4}
\end{figure}

\subsection{The scaling exponent $\beta$}
In the main paper we show that our data is consistent with $\beta =1/2$. Here we provide additional analysis to confirm this.

First, at long times $\ell^{1/\beta}(t)$ is $\propto t - t ^*$ only for the correct $\beta$. We perform such linear fits for all our $13$ data sets (see Fig.~\ref{fig:3}) assuming different values of $\beta$, and in Extended Data Fig.~\ref{fig:S3}{\ba} show the combined $\chi^2$ values for all $13$ fits, which show a minimum at $1/\beta \approx2$.

\begin{figure*}[t!]
\centerline{\includegraphics[width=\textwidth]{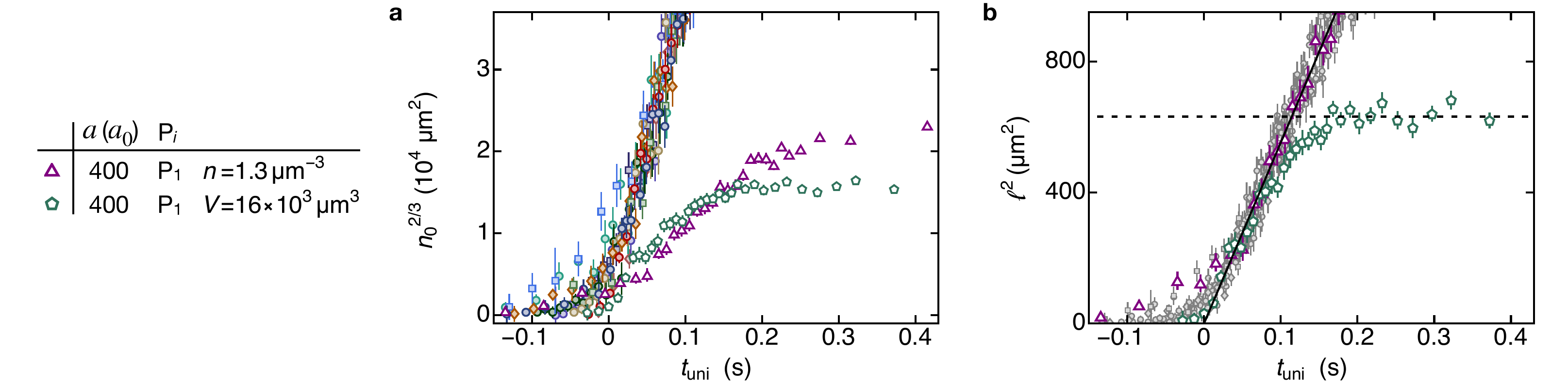}}
\caption{{\textbf{Changing $n$ or $V$.}
 {\ba}, Plotting $n_0^{2/3}$ versus $\tuni = t - t^*$ for all $11$ data sets taken with the same $n$ and $V$ (see Extended Data Fig.~\ref{fig:S2}) together with the two additional sets taken with reduced $n$ (purple triangles) or $V$ (green pentagons) shows that the rate of growth of $n_0^{2/3}$ depends on the gas density and the system size. {\bb}, However, the rate of growth of $\ell^2$ in the scaling regime is universal. The only effect of the system size is that for lower $V$ the saturation of $\ell^2$ occurs at a lower value (dashed line shows $V^{2/3}$ for the smaller box).}
}
\label{fig:S5}
\end{figure*}

Second, ignoring the linear-fit quality for different $\beta$ values, for self-consistency $\upd \ell^{1/\beta}/ \upd t$ should be $\propto (na)^{1-1/(2\beta)}$ [Eq.~(\ref{eq:Ddim})], so $C = (na)^{1/(2\beta)-1}\upd \ell^{1/\beta} /\upd t $ should be independent of $na$. Fitting $C \propto (na)^{\gamma}$ for all $13$ data sets, we obtain $\gamma (\beta)$ shown in Extended Data Fig.~\ref{fig:S3}{\bb}. We see that the results are self-consistent ($\gamma = 0$ within errors) only for $\beta = 0.50(3)$.

{\subsection{Dynamic scaling of $n_k$}

As a complement to Fig.~\ref{fig:1}{\bd}, in Extended Data Fig.~\ref{fig:S4} (bottom panels) we show the log-space spread of the data around their geometric mean at each $k$, both before (left) and after (right) the dynamic scaling.
}

{For this dynamical collapse we do not attempt to account for the effects of our finite $k$-space resolution (see Methods section `Deducing $\ell$ from $n_0$'), which would involve numerically deconvolving the observed $n_k(k)$ distributions with a model point spread function. For this data, $\ell/\ell_0<0.5$ and most points lie at $k\gg\sigma_k$, so the resolution effects should be small. Numerically convolving model functions similar to our $n_k$ and a Gaussian with our $\sigma_k$, we indeed find that the effects on the $n_k$ values are of the order of a few \%, significantly smaller than the residual spread in the bottom right panel of Extended Data Fig.~\ref{fig:S4}.}

\subsection{Changing $n$ or $V$}

For two of the measurements summarised in Fig.~\ref{fig:3} we reduced either $n$ or $V$ (while keeping $\varepsilon$ the same).
{In Extended Data Fig.~\ref{fig:S5} we show further details of these measurements.
Reducing either $n$ or $V$ reduces the equilibrium value of $n_0$ and the rate of growth of $n_0^{2/3}$ in the scaling regime, but the scaling dynamics of $\ell^2$ are universal.
}

\subsection{The time constant $\tau$ for the approach to the scaling regime}

\begin{figure}[b]
\centerline{\includegraphics[width=\columnwidth]{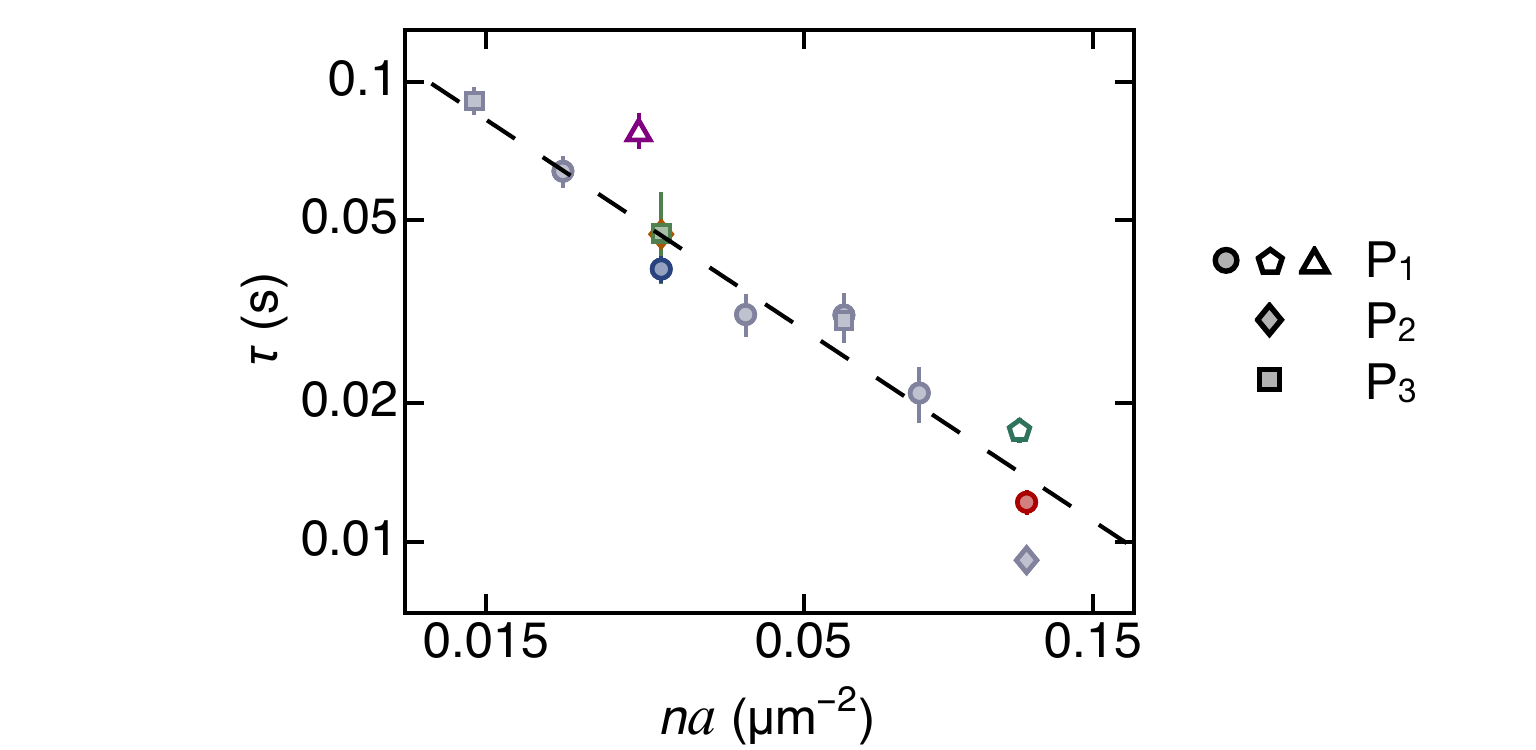}}
\caption{\textbf{The timescale for the approach to the scaling regime.}
The timescale $\tau$ of the exponential fits to the data before the scaling regime, for all $13$ data sets summarised in Fig.~\ref{fig:3}. The dashed line is a power law fit, giving $\tau\propto (na)^{-0.9(1)}$, consistent with $\tau \propto t_\xi \propto 1/(na)$.
}
\label{fig:S6}
\end{figure}

In Extended Data Fig.~\ref{fig:S6}, we show the results of independent exponential fits to the early-time data (the approach to the universal scaling regime) for all $13$ sets summarised in Fig.~\ref{fig:3}.
Fitting $\tau \propto (na)^\delta$ with a free exponent $\delta$ (dashed line) gives $\delta = - 0.9(1)$, consistent with $\tau \propto t_\xi \propto 1/(na)$. 

{\subsection{Particle loss}
The one-body particle loss due to the background gas in the vacuum chamber and off-resonant light scattering is always small (a few \%) during $1$\,s of gas relaxation (see Extended Data Fig.~\ref{fig:S2}), but the loss due to three-body recombination, $-{\rm d}n/{\rm d}t\propto n^3a^4$, grows with $n$ and $a$, and limits the range of interaction strengths we can explore. For our larger $n=5.4(1.2)\upmu$m$^{-3}$ and largest $a=400a_0$ the total particle loss over $1$\,s is just over $10\%$; for all other measurements it is much lower.
}

\begin{figure}[b!]
\centerline{\includegraphics[width=\columnwidth]{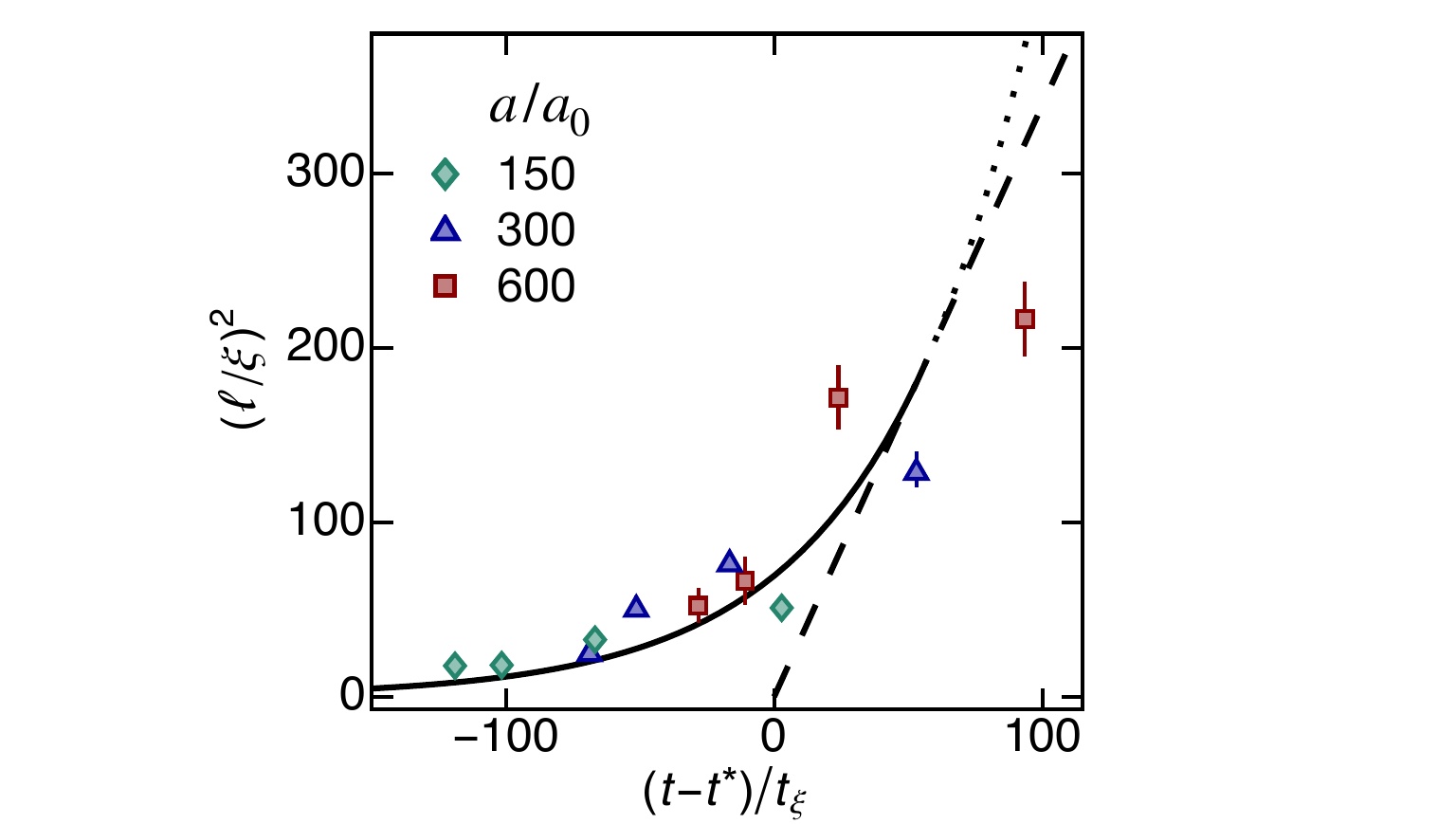}}
\caption{\textbf{Comparison to the data from Ref.~\cite{Glidden:2021}.}
We present the data corresponding to Fig.~3 in Ref.~\cite{Glidden:2021} in the style of our Fig.~\ref{fig:4}{\bb}, and show that they are consistent with our results if we include appropriate time shifts $t^*$ (which were not considered in Ref.~\cite{Glidden:2021}). The solid and dashed lines are the same as in Fig.~\ref{fig:4}, and the dotted one is the extension of the former. For the largest $a$, the data reaches $(\ell/\xi)^2 \approx 200$, at which point it is equally consistent with the speed limit (dashed line) and with the extrapolation of the initial exponential growth (dotted line).
}
\label{fig:S7}
\end{figure}

\subsection{Comparison to previous measurements}

The early experiments on condensation dynamics were performed with inhomogeneous gases in harmonic traps~\cite{Miesner:1998b, Kohl:2002, Ritter:2007, Hugbart:2007,Smith:2012}, which makes comparison with uniform-system theory difficult. 
The closest one to probing the uniform-system physics was Ref.~\cite{Ritter:2007}, where the emergence of coherence in a $^{87}$Rb gas was probed interferometrically and the authors focused on the quasi-homogeneous region near the trap centre. Still, for a quantitative comparison, a problem is that in a harmonic trap the central density grows, and hence $\xi$ decreases, during condensation. We therefore make only a rough comparison to our results. In Ref.~\cite{Ritter:2007}, coherence spread over $8.5\,\um$ in about $350\,\textrm{ms}$. 
However, most of the dynamics happened over about $150\,\textrm{ms}$, which gives an estimate $\upd \ell^2/\upd t \approx 0.5\,\um^2/\textrm{ms}$, about five times lower than $3.4\, \hbar/m_{\rm Rb} \approx 2.5\,\um^2/\textrm{ms}$.

In Ref.~\cite{Glidden:2021} (by our group), relaxation was studied in a box trap and for tunable interactions. The box size and the range of $a$ values were similar to ours, but the initial far-from-equilibrium state was prepared by rapid evaporation, removing $77\%$ of the atoms, so the gas density and the observable values of $\ell/\xi$ were lower. The data were within errors consistent with dynamic scaling, but the deduced $\beta \approx 0.34$ was slightly lower than the expected $1/2$, and the characteristic relaxation time was $\propto 1/a$. However, as with all the NTFP experiments prior to Ref.~\cite{Gazo:2025}, the approximation $|t^*| \ll t$ was implicitly assumed. 
In Extended Data Fig.~\ref{fig:S7}, we show that these measurements can be aligned with our results in Fig.~\ref{fig:4}{\bb} simply by including appropriate time shifts $t^*$, and the observed $a$-dependence of the relaxation dynamics is explained by the system not having reached the universal speed limit; all the data are consistent with the initial exponential growth that has a characteristic timescale 
$\tau = 56 \, t_{\xi}\propto 1/(na)$.

\end{document}